# FNHSM_HRS: Hybrid recommender system using fuzzy clustering and heuristic similarity measure

Mostafa Khalaji*, Faculty of Computer Engineering, K. N. Toosi University of Technology, Tehran, Iran, Email: Khalaji@email.kntu.ac.ir, ORCID: 0000-0002-5019-1824

Chitra Dadkhah, Faculty of Computer Engineering, K. N. Toosi University of Technology, Tehran, Iran

**Abstract**

Nowadays, Recommender Systems have become a comprehensive system for helping and guiding users in a huge amount of data on the Internet. Collaborative Filtering offers to active users based on the rating of a set of users. One of the simplest and most comprehensible and successful models is to find users with a taste in recommender systems. In this model, with increasing number of users and items, the system is faced to scalability problem. On the other hand, improving system performance when there is little information available from ratings, that is important. In this paper, a hybrid recommender system called FNHSM_HRS which is based on the new heuristic similarity measure (NHSM) along with a fuzzy clustering is presented. Using the fuzzy clustering method in the proposed system improves the scalability problem and increases the accuracy of system recommendations. The proposed system is based on the collaborative filtering model and is partnered with the heuristic similarity measure to improve the system's performance and accuracy. The evaluation of the proposed system based results on the MovieLens dataset carried out the results using MAE, Recall, Precision and Accuracy measures Indicating improvement in system performance and increasing the accuracy of recommendation to collaborative filtering methods which use other measures to find similarities.

---

**چکیده:** امروزه سیستم‌های توصیه‌گر به یک سیستم فراگیر برای راهنمایی و هدایت کاربران در حجم عظیمی از داده در اینترنت، تبدیل شده است. پالایش همکارانه که پیشنهادات به کاربر فعال را براساس امتیازدهی مجموعه‌ای از کاربران ارائه می‌دهد، یکی از مدل‌های ساده و قابل درک و موفق برای پیدا کردن افراد هم سلیقه در سیستم‌های توصیه‌گر است. در این مدل، با افزایش تعداد کاربران و اقلام، سیستم دچار مشکل مقیاس پذیری می‌شود. از طرفی دیگر بهبود عملکرد سیستم در مواقعی که اطلاعات کمی از امتیازات دردسترس داریم، امری مهم است. در این مقاله یک سیستم توصیه‌گر ترکیبی به نام FNHSM_HRS که مبتنی بر معیار شباهت اکتشافی (NHSM) به همراه خوشه‌بندی فازی است، ارائه شده است. استفاده از روش خوشه‌بندی فازی در سیستم پیشنهادی باعث بهبود مساله مقیاس پذیری گشته و دقت پیشنهادات سیستم را افزایش می‌دهد. سیستم پیشنهادی مبتنی بر مدل پالایش همکارانه بوده و با استفاده از معیار شباهت اکتشافی، عملکرد و صحت سیستم را ارتقا می دهد. ارزیابی نتایج سیستم پیشنهادی برروی مجموعه داده MovieLens صورت گرفته، نتایج ارزیابی با استفاده از معیارهای MAE، Accuracy، Precision و Recall بیانگر بهبود کارایی سیستم و افزایش دقت پیشنهادات نسبت به روش‌های پالایش همکارانه‌ای که از معیارهای دیگری برای پیدا کردن شباهت استفاده می نمایند، می‌باشد.

**واژه های کلیدی:** سیستم‌های توصیه‌گر، پالایش همکارانه، خوشه‌بندی فازی، معیار شباهت اکتشافی

## ۱. مقدمه

با افزایش اطلاعات در اینترنت و فضای مجازی و خریدهای اینترنتی و تعاملات کاربران با یکدیگر، سیستم‌های توصیه‌گر(Recommender Systems) جهت هدایت کاربران به سمت سلایق یا نیازهایی که دارند، در بیست سال اخیر و به ویژه در دهه‌ی اول قرن بیست و یکم مورد مطالعه قرار گرفته‌اند و پژوهش‌های بسیاری در این زمینه انجام شده است [۱]. سیستم‌های توصیه‌گر براساس نحوه پیشنهاد دهی به مدل‌های مختلفی دسته بندی می‌شوند که یکی از مهم ترین آنها مبتنی بر پالایش همکارانه(Collaborative Filtering) است. عملکرد این روش به این طریق است که براساس شباهت بین کاربران یا اقلام، پیش بینی و پیشنهاد دهی به کاربر فعال را انجام می‌دهد. از این رو شامل تکنیک‌های مبتنی بر حافظه و مبتنی بر مدل است [۲]. مدل مبتنی بر حافظه در ابتدا شباهت بین کاربران را محاسبه و پس از آن کاربران مشابه به کاربر فعال را به عنوان کاربران همسایه انتخاب کرده و در پایان پیشنهادات را براساس این کاربران به کاربر فعال ارائه می‌کند.

پالایش همکارانه مبتنی بر مدل در ابتدا یک مدلی از رفتار کاربران را ایجاد می‌کند و پس از آن امتیازات اقلام مشاهده نشده را براساس آن مدل پیش بینی می نماید. شروع سرد(Cold Start) و مقیاس پذیری(Scalability) مشکلاتی در سیستم‌های توصیه‌گر هستند. هنگامی که سیستم توصیه‌گر با کمبود اطلاعات(امتیازات) از سوی سوابق کاربر مواجه می‌شود، مشکل شروع سرد از نوع کاربر جدید در سیستم بوجود می‌آید و در صورتیکه قلم جدیدی وارد سیستم شود از آنجایی‌که کاربران سیستم این قلم را قبلا مشاهده ننموده‌اند، مشکل شروع سرد برای قلم جدید رخ خواهد داد. از سوی دیگر با افزایش چشمگیر تعداد کاربران در فضای مجازی، مشکل مقیاس پذیری برای سیستم‌های توصیه‌گر رخ می‌دهد که می‌توان گفت سیستم عملکرد و دقت خود را تا حدی از دست خواهد داد. سیستم توصیه‌گر از نوع پالایش همکارانه مدل‌های بسیار ساده و قابل فهم و به راحتی قابل پیاده سازی می‌باشند.

در این مقاله تمرکز اصلی سیستم توصیه‌گر پیشنهادی بر روی عملکرد سیستم با استفاده از ترکیبی از روش‌های مبتنی بر حافظه و مبتنی بر مدل جهت رفع مشکل مقیاس پذیری است. هسته اصلی

---

* **نویسنده مسئول**

آدرس پست الکترونیک: Khalaji@email.kntu.ac.ir

این سیستم در ابتدا از روش خوشه‌بندی فازی[2] و الگوریتم‌های پالایش همکارانه مشتق شده است که از معیار شباهت اکتشافی که در مرجع [۳] با نام NHSM[3] معرفی شده است، جهت تعیین کاربران همسایه با کاربر فعال استفاده می نماید. استفاده از روش‌های خوشه‌بندی در این نوع سیستم‌های توصیه‌گر باعث می‌شود تا کاربرانی که از نظر سلیقه به یکدیگر دارای شباهت‌هایی هستند در گروه‌های خاصی خوشه‌بندی شوند و پس از آن عملیات پیش بینی و پیشنهاد دهی اقلام در هر خوشه به صورت مجزا صورت گیرد و همین امر باعث بهبود عملکرد سیستم توصیه‌گر از نظر زمان و دقت می‌شود.

ساختار مقاله به این صورت می‌باشد که در بخش دوم به مروری برکارهای محققین می پردازیم. در بخش سوم سیستم توصیه‌گر پیشنهادی FNHSM_HRS[4] معرفی می‌شود. بخش چهارم، بخش ارزیابی و نتایج پیاده سازی می‌باشد که در آن به نتایج آزمایشات و مقایسه با روش‌های دیگر پرداخته می‌شود. در نهایت در بخش آخر نیز نتیجه گیری ارائه می‌گردد.

## ۲. مروری بر کارهای محققین

سیستم‌های توصیه‌گر در ابتدا توسط Goldberg et.al معرفی شدند [۴]. پس از آن چندین روش براساس نحوه پیشنهاد دهی ارائه شدند که یکی از آنها روش پالایش همکارانه است. پالایش همکارانه به عنوان یکی از محبوب ترین روش پیشنهاد دهی خصوصی سازی شده برای کاربران است که در بسیاری از حوزه‌ها مورد استفاده قرار می گیرد. این روش براساس یک سری معیارهای شباهت و یکسری مدل‌های از پیش تعریف شده، عملیات خود را انجام می‌دهند. اگرچه این روش از مشکلاتی از قبیل شروع سرد، پراکندگی داده و مقیاس پذیری رنج می‌برد اما فهم و پیاده سازی آنها بسیار راحت است و از مدل‌های پایه در سیستم‌های توصیه‌گر می باشند.

برای بهبود عملکرد سیستم، بسیاری از پژوهشگران انواع روش‌های معیار شباهت را معرفی کرده‌اند. این معیارها شامل معیار شباهت Pearson، Cosine، PIP و غیره [۵ و ۶] است.

Bellogin et.al روش‌هایی را برای بهبود عملکرد سیستم‌های توصیه‌گر معرفی کردند که از روش‌های Herlocker's weighting و McLaughlin's weighting جهت تعیین کاربرانی که از نظر سلیقه به کاربر فعال نزدیک بودند را انتخاب می‌کرد [۷]. همچنین Haifeng et.al یک معیار شباهت اکتشافی با نام NHSM ارائه کردند که سه جنبه مجاورت[5]، اثر شدید[6] و محبوبیت[7] امتیازات کاربران را در هنگام انتخاب کاربران همسایه برای کاربر فعال مد نظر داشت. از سوی دیگر یک معیار ترکیب شده از Jaccard و Mean Squared Difference توسط Bobadila et.al و همکارانش ارائه گردید [۸].

بکارگیری روش‌های خوشه‌بندی برای گروه بندی کاربران مشابه از نظر سلیقه کمک شایانی به حل مسائل مقیاس پذیری می‌کند. از این رو Koohi et.al با بکارگیری خوشه‌بندی فازی و روش فازی‌زدایی[8] Max، به صورت تخصیص کاربران به تمام خوشه‌ها با درجه عضویت متفاوت و استفاده از  معیار شباهت Pearson برای پیداکردن نزدیک ترین همسایه، نشان دادند که عملکرد سیستم آنها نسبت به استفاده از روش‌های K-Means و SOM بهبود بخشیده شده است [۹].

## ۳. سیستم FNHSM_HRS

در شکل شماره ۱ ساختار سیستم پیشنهادی FNHSM_HRS نشان داده شده است. سیستم FNHSM_HRS دارای دو بخش برون خط و برخط است که در بخش برون خط مدل سیستم براساس اطلاعات داخل ماتریس امتیاز دهی کاربر-قلم آموزش داده می‌شود. در بخش بر خط پیشنهادات سیستم براساس مدل بدست آمده در مرحله برون خط برای کاربر فعال ارائه می‌گردد.

### ۳.۱. بخش برون خط

روش‌های سنتی سیستم‌های توصیه‌گر، تنها به پیدا کردن کاربرانی که با کاربر فعال هم سلیقه هستند به انواع روش‌های معیار شباهت، اکتفا می‌کردند که برای مواقعی که ماتریس امتیازات کاربران، تنک[9] است، کارا نمی‌باشد. از سوی دیگر اکثر این معیارها از مشکل

---

[2] Fuzzy C-Means
[3] New Heuristic Similarity Measure
[4] Fuzzy NHSM Hybrid Recommender System
[5] Proximity
[6] Impact
[7] Popularity
[8] Defuzzification
[9] Sparse

پیچیدگی زمانی رنج می‌برند. برای حل این مشکلات، با خوشه‌بندی کردن کاربران در خوشه‌های متفاوت در سیستم پیشنهادی ، باعث تسریع در روند عملکرد سیستم شده و بیانگر این مطلب می‌باشد که کاربران موجود در هرخوشه از نظر سلیقه باهم دارای شباهت‌هایی هستند.

در این بخش ماتریس امتیازات به عنوان ورودی به سیستم توصیه‌گر داده شده است؛ ۸۰ درصد ماتریس امتیازات به عنوان مجموعه داده آموزشی انتخاب و با استفاده از روش خوشه‌بندی فازی کاربران سیستم به تعدادی خوشه تقسیم می‌شوند. در سیستم FNHSM_HRS با توجه به پیشنهادات محققین[۹]، ۳ خوشه در نظر گرفته شده است. الگوریتم خوشه‌بندی فازی با تخصیص درجه عضویت متفاوت به کاربران در هر خوشه، تاثیر کاربران در هر خوشه را مشخص می‌نماید. برای رتبه بندی درجه کاربران و انتخاب کاربران در مرحله پیش بینی سیستم از روش فازی زدایی درجات مبتنی بر مرکز ثقل (COG)[۱۰] استفاده شد.

### ۳.۲. بخش بر خط

در این بخش، کاربر فعال که قرار است سیستم پیشنهاداتی در رابطه با فیلم‌های مشاهده نشده به او ارائه نماید، وارد سیستم شده و براساس سوابق خود کاربر فعال، عملیات محاسبات نزدیک ترین همسایه در خوشه کاربر فعال توسط معیار شباهت NHSM صورت می‌گیرد. این معیار دارای دو ضریب اصلی است که در رابطه (۱) مشخص شده است.

$$NHSM\_Sim(u,v) = JPSS\_Sim(u,v) \cdot URP_{Sim(u,v)} \quad (1)$$

برای محاسبه معیار شباهت $NHSM\_Sim$ باید در ابتدا معیار شباهت $JPSS\_Sim$ محاسبه شود که خود نیز از دو معیار شباهت مشتق شده است که در روابط (۲) و (۳) ذکر شده است.

$$JPSS\_Sim(u,v) = PSS\_Sim(u,v) \cdot Jaccard'_{Sim(u,v)} \quad (2)$$

$$Jaccard'_{Sim(u,v)} = \frac{|I_u \cap I_v|}{|I_u| \times |I_v|} \quad (3)$$

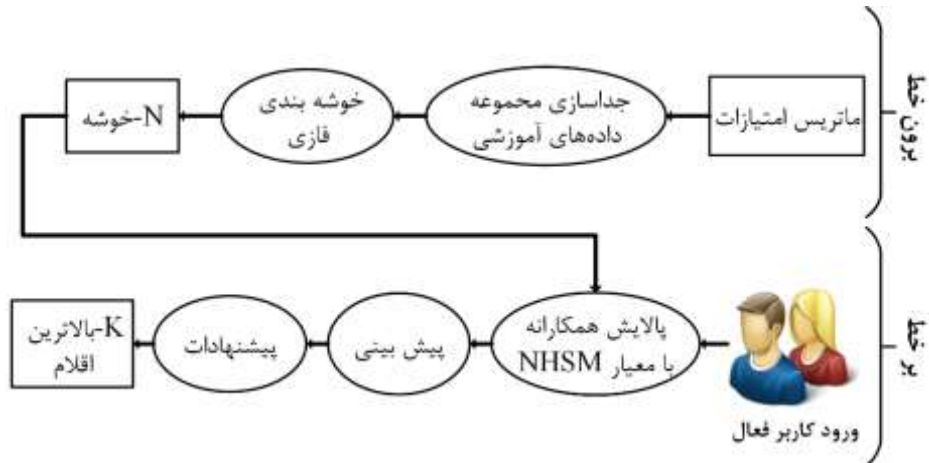

شکل شماره ۱: سیستم پیشنهادی FNHSM_HRS

$|I_u \cap I_v|$ بیانگر تعداد اقلام مشابه ای است که کاربر u و v مشاهده نموده اند و $|I_u|$ بیانگر تعداد اقلامی که کاربر فعال (u) امتیاز داده است و $|I_v|$ بیانگر تعداد اقلامی که کاربر دیگر (v) امتیاز داده است. معیار $PSS\_Sim$ از طریق رابطه (۴) بدست می‌آید.

$$PSS_{Sim}(r_{u,p}, r_{v,p}) = Proximity(r_{u,p}, r_{v,p}) \cdot Significance(r_{u,p}, r_{v,p}) \cdot Singularity(r_{u,p}, r_{v,p}) \quad (4)$$

$Proximity$، $Significance$ و $Singularity$ از طریق روابط (۵) و (۶) و (۷) بدست می‌آید.

$$Proximity(r_{u,p}, r_{v,p}) = 1 - \frac{1}{1+\exp(-|r_{u,p} - r_{v,p}|)} \quad (5)$$

$$Significance(r_{u,p}, r_{v,p}) = \frac{1}{1+\exp(-|r_{u,p} - r_{med}| \cdot |r_{v,p} - r_{med}|)} \quad (6)$$

$$Singularity(r_{u,p}, r_{v,p}) = 1 - \frac{1}{1+\exp(-|\frac{r_{u,p}+r_{v,p}}{2} - \mu_p|)} \quad (7)$$

$r_{u,p}$ امتیاز قلم p توسط کاربرفعال u و $r_{v,p}$ امتیاز قلم p توسط کاربر v است. $r_{med}$ هم میانگین امتیازاتی بین امتیازاتی که کاربر موظف است برای امتیازدهی به یک قلم وارد کند، می‌باشد. ماتریس امتیازدهی در سیستم توصیه‌گر FNHSM_HRS دارای محدوده امتیازات بین عدد ۱ تا ۵ است که متوسط آن ۳ می‌باشد. $\mu_p$ هم میانگین امتیازات قلم p توسط کاربران است.

---

[۱۰] Center of Gravity

آخرین مرحله از رابطه (۱)، محاسبه معیار شباهت URP_Sim است که با استفاده از رابطه (۸) بدست می‌آید.

$$URP_{Sim(r_{u,p},r_{v,p})} = 1 - \frac{1}{1 + \exp(-|\mu_u - \mu_v|.|\sigma_u - \sigma_v|)} \quad (8)$$

$\mu_u$ میانگین امتیازات و $\sigma_u$ انحراف معیار کاربر فعال است که طبق رابطه (۹) بدست می‌آید.

$$\sigma_u = \sqrt{\sum_{p\epsilon I_u}(r_{u,p} - \bar{r_u})^2/|I_u|} \quad (9)$$

در مرحله بعدی تعداد کاربران همسایه براساس بالاترین درجه شباهت کاربران همسایه به کاربر فعال در خوشه مربوطه، مشخص و انتخاب می‌شود. از این رو با توجه به رابطه (۱۰) عملیات پیش بینی امتیازات اقلام مشاهده نشده براساس پالایش همکارانه در سیستم FNHSM_HRS، محاسبه می‌شود.

$$Predict(u,i) = \mu_u + \frac{\sum_{j=1}^{m}(r_{v_j,i} - \mu_v).NHSM\_Sim(u,v_j)}{\sum_{j=1}^{m}|NHSM\_Sim(u,v_j)|} \quad (10)$$

در اینجا، $u$ کاربر فعال و $i$ فیلمی است که سیستم FNHSM_HRS قرار است برای آن، امتیازی را پیش بینی نماید. سیستم FNHSM_HRS برای کلیه اقلام مشاهده نشده توسط کاربر فعال این امتیازات را محاسبه نموده و k قلم برتر را پیشنهاد می دهد.

در رابطه ۱۰، $\mu_u$، میانگین امتیازات کاربر فعال و $m$ تعداد کاربران همسایه داخل خوشه‌ی کاربر فعال است. $NHSM\_Sim(u,v_j)$، میزان شباهت بدست آمده کاربر فعال u با کاربر $v_j$ است. $r_{v,i}$، امتیاز کاربر v به فیلم $i$ است و $\mu_v$، میانگین امتیازات کاربرv است.

## ۴. ارزیابی سیستم FNHSM_HRS

داده‌های ورودی سیستم FNHSM_HRS مجموعه داده MovieLens شامل ۹۴۳ کاربر و ۱۶۸۲ فیلم با ۱۰۰ هزار تا امتیاز کاربر به فیلم‌ها، در نظر گرفته شده است [۱۰]. بازه امتیاز دهی در این مجموعه داده از ۱ تا ۵ است که به ترتیب یک نشان دهنده مورد پسند نبودن و عدد ۵ نشان دهنده مورد پسند بودن کاربر به یک فیلم خاص میباشد.

برای ارزیابی عملکرد سیستم، از 5-fold مجموعه داده‌ها به روش Cross-validation، استفاده شده است که ۸۰ درصد داده‌ها برای آموزش و ایجاد مدل سیستم پیشنهادی و ۲۰ درصد داده‌ها برای آزمایش سیستم در نظر گرفته شده است. ارزیابی سیستم براساس معیارهای MAE،

Accuracy، Precision و Recall طبق رابطه‌های (۱۱-۱۴) برروی داده‌های آزمایش محاسبه شده است که در جدول شماره ۱ ماتریس اغتشاش[11] مربوط را مشاهده می‌نمایید [۱۱].

| Actual / Predicted | Negative | Positive |
|---|---|---|
| Negative | A | B |
| Positive | C | D |

جدول شماره ۱: ماتریس اغتشاش [۱۱]

$$MAE = \frac{\sum_{i=1}^{n}|\hat{r}_{u,i} - r_{u,i}|}{n} \quad (11)$$

$$Accuracy = \frac{Correct\ Recommendation}{Total\ Possible\ Recommedation} = \frac{A+D}{A+B+C+D} \quad (12)$$

$$Precision = \frac{Correctly\ Recommended\ Items}{Total\ Recommeded\ Items} = \frac{D}{B+D} \quad (13)$$

$$Recall = \frac{Correctly\ Recommended\ Items}{Total\ Useful\ Recommedations} = \frac{D}{C+D} \quad (14)$$

همانطور که قبلا اشاره نموده‌ایم در سیستم FNHSM_HRS تعداد خوشه‌ها را ۳ و تعداد کاربران همسایه را ۵۰ در نظر گرفتیم. سیستم ۵ بار مستقل برای هر Top-N جداگانه اجرا و نتایج ذخیره گردیده است. در سیستم FNHSM_HRS از معیار شباهت اکتشافی استفاده شده است که در نتیجه با ۵ روش دیگری: روش Pearson، روش Cosine، روش وزن‌دار McLaughlin's weighting [۷]، روش وزن‌دار Herlocker's weighting [۷] و روش RA [۱۲] مقایسه و ارزیابی شده است که نتایج را در جداول شماره (۲) و شماره (۳) مشاهده می نمایید.

---

[11] Confusion Matrix

<gnoreimage_ref id="1" />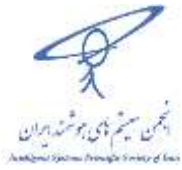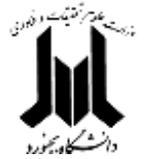

هفتمین کنگره مشترک سیستمهای فازی و هوشمند ایران
هجدهمین کنفرانس سیستمهای فازی و هفدهمین کنفرانس سیستمهای هوشمند
9-11 بهمن 1397    بجنورد، ایران

| نام روش | معیار ارزیابی | 5 بالاترین اقلام | 10 بالاترین اقلام | 15 بالاترین اقلام | 20 بالاترین اقلام | 30 بالاترین اقلام |
|---|---|---|---|---|---|---|
| FNHSM_HRS | Accuracy | 60.203 | 59.373 | 56.200 | 55.557 | 55.155 |
|  | Precision | 91.391 | 91.148 | 90.962 | 90.902 | 90.844 |
|  | Recall | 58.643 | 51.029 | 47.560 | 45.486 | 44.379 |
|  | MAE | 0.756 | 0.766 | 0.778 | 0.782 | 0.789 |
| F_CF with Pearson | Accuracy | 59.537 | 56.756 | 55.784 | 55.160 | 54.762 |
|  | Precision | 90.674 | 90.352 | 90.230 | 90.147 | 90.082 |
|  | Recall | 54.537 | 50.090 | 46.790 | 44.834 | 42.768 |
|  | MAE | 0.768 | 0.775 | 0.782 | 0.787 | 0.794 |
| F_CF with Cosine | Accuracy | 60.010 | 57.275 | 56.035 | 55.073 | 53.610 |
|  | Precision | 82.020 | 81.371 | 80.935 | 80.683 | 80.441 |
|  | Recall | 50.278 | 50.206 | 47.041 | 45.421 | 44.003 |
|  | MAE | 0.827 | 0.828 | 0.829 | 0.834 | 0.839 |
| F_MW | Accuracy | 59.507 | 57.717 | 56.113 | 55.734 | 55.056 |
|  | Precision | 91.098 | 90.856 | 90.691 | 90.617 | 90.550 |
|  | Recall | 58.039 | 51.015 | 47.576 | 45.433 | 43.383 |
|  | MAE | 0.758 | 0.768 | 0.778 | 0.783 | 0.791 |
| F_HW | Accuracy | 58.172 | 57.002 | 55.929 | 55.406 | 54.904 |
|  | Precision | 90.881 | 90.562 | 90.500 | 90.429 | 90.364 |
|  | Recall | 58.220 | 50.604 | 47.074 | 45.136 | 43.042 |
|  | MAE | 0.777 | 0.785 | 0.791 | 0.796 | 0.803 |
| F_RA | Accuracy | 59.887 | 58.674 | 56.289 | 54.601 | 54.035 |
|  | Precision | 89.650 | 89.355 | 89.209 | 89.150 | 89.088 |
|  | Recall | 57.513 | 50.297 | 49.554 | 44.385 | 44.067 |
|  | MAE | 0.760 | 0.768 | 0.777 | 0.783 | 0.791 |

جدول شماره 2: نتایج ارزیابی سیستم FNHSM_HRS

| نام روش | معیار ارزیابی | میانگین بالاترین n ها | نام روش | معیار ارزیابی | میانگین بالاترین n ها |
|---|---|---|---|---|---|
| FNHSM_HRS | Accuracy | 57.2976 | F_MW | Accuracy | 56.8254 |
|  | Precision | 91.0494 |  | Precision | 90.7624 |
|  | Recall | 49.4194 |  | Recall | 49.0892 |
|  | MAE | 0.7742 |  | MAE | 0.7756 |
| F_CF with Pearson | Accuracy | 56.3998 | F_HW | Accuracy | 56.2826 |
|  | Precision | 90.297 |  | Precision | 90.5472 |
|  | Recall | 47.8038 |  | Recall | 48.8152 |
|  | MAE | 0.7812 |  | MAE | 0.7904 |
| F_CF with Cosine | Accuracy | 56.4006 | F_RA | Accuracy | 56.6972 |
|  | Precision | 81.09 |  | Precision | 89.2904 |
|  | Recall | 47.3898 |  | Recall | 49.1632 |
|  | MAE | 0.8312 |  | MAE | 0.7758 |

جدول شماره 3: میانگین Top-N برای هر روش

## ۵. نتیجه گیری

هدف اصلی سیستم‌های توصیه‌گر ارائه یک سری پیشنهادات براساس سلیقه کاربر و پیداکردن کاربرانی که از نظر سلیقه با کاربر فعال دارای شباهت‌های زیادی است، می‌باشد. از این رو یکی از چالش‌های اصلی این سیستم‌ها دقت و عملکرد آنها با توجه به حجم زیاد اطلاعات(افزایش تعداد کاربران و اقلام سیستم) در کمترین زمان ممکن است. روش ارائه شده در سیستم FNHSM_HRS که حاصل ترکیب روش‌های خوشه‌بندی فازی و بکارگیری معیار شباهت NHSM است، کمک به سزایی به بهبود Accuracy، Precision، Recall و MAE کرده است. نتایج آزمایشات سیستم پیشنهادی FNHSM_HRS به صورت میانگین، MAE برابر با ۰٫۷۷۴۲، Accuracy برابر با ۵۷٫۲۹۷۶، Precision برابر با ۹۱٫۰۴۹۴ و Recall برابر با ۴۹٫۴۱۹۴ شده است که در مقایسه با سایر معیارهای شباهت، بیانگر بهبود عملکرد سیستم و افزایش دقت می‌باشد. سیستمFNHSM_HRS مشکل مقیاس پذیری در سیستم‌های توصیه‌گر را به کمک روش خوشه‌بندی حل نموده است.

## ۶. مراجع

[1] C. Aggarwal, *Recommender systems*. Springer International Publishing, (2016).

[2] F. Cacheda, V. Carneiro, D. Fernández and V. Formoso, Comparison of collaborative filtering algorithms: Limitations of current techniques and proposals for scalable, high-performance recommender systems, *ACM Transactions on the Web (TWEB),* **5.1**, (2011).

[3] H. Liu, Z. Hu, A. Mian, H. Tian and X. Zhu, A new user similarity model to improve the accuracy of collaborative filtering, Knowledge-Based Systems, **56** (2014),  156-166.

[4] D. Goldberg, D. Nichols, B. Oki and D. Terry, Using collaborative filtering to weave an information tapestry, *Communications of the ACM,* **35.12**, (1992), 61-70.

[5] H. Ahn, A new similarity measure for collaborative filtering to alleviate the new user cold-starting Problem, *Information Sciences,* **178.1**, (2008),  37-51.

[6] G. Adomavicius and A. Tuzhilin, Toward the next generation of recommender systems: A survey of the state-of-the-art and possible extensions, *IEEE Transactions on Knowledge & Data Engineering,* **6**, (2005), 734-749.

[7] A. Bellogín, P. Castells and I. Cantador, Improving memory-based collaborative filtering by neighbour selection based on user preference overlap, *Proceedings of the 10th Conference on Open Research Areas in Information Retrieval*, (2013).

[8] J. Bobadilla, F. Ortega, A. Hernando and J. Bernal, A collaborative filtering a roach to mitigate the new user cold start problem, *Knowledge-Based Systems,* **26**, (2012),  225-238.

[9]  H. Koohi and K. Kiani, User based Collaborative Filtering using fuzzy C-means, *Measurement,* **91**, (2016), 134-139.

[10] J. Konstan, B. Miller, D. Maltz, J. Herlocker, L. Gordon and J. Riedl, GroupLens: a lying collaborative filtering to Usenet news, *Communications of the ACM,* **40.3**, (1997), 77-87.

[11] M. Hahsler, Developing and testing top-n recommendation algorithms for 0-1 data using recommenderlab, *NSF Industry University Cooperative Research Center for Net-Centric Software and System,* (2011).

[12]  A. Javari, J. Gharibshah and M. Jalili, Recommender systems based on collaborative filtering and resource allocation, *Social Network Analysis and Mining,* **4.1**, (2014).